\documentclass[epsfig,12pt,onecolumn]{article}
\usepackage{amsfonts}
\usepackage{amssymb}
\usepackage{amsmath}
\usepackage{multicol}
\usepackage{graphicx}
\usepackage{float}
\usepackage{caption}

\input{tcilatex}
\makeatletter \@addtoreset{equation}{section}

\def \be{\begin{equation}}
\def \ee{\end{equation}}
\def \bea{\begin{eqnarray}}
\def \eea{\end{eqnarray}}

\newcommand{\nc}{\newcommand}
\nc{\al}{\alpha} \nc{\bib}{\bibitem} \nc{\la}{\lambda}
\nc{\C}{\mbox{\hspace{1.24mm}\rule{0.2mm}{2.5mm}\hspace{-2.7mm} C}}
\nc{\R}{\mbox{\hspace{.04mm}\rule{0.2mm}{2.8mm}\hspace{-1.5mm} R}}

\begin{document}

\title{\textbf{Cosmic acceleration in Lovelock quantum gravity}}
\author{M. Bousder$^{1}$ \thanks{%
mostafa.bousder@gmail.com}, A.Riadsolh$^{3}$\thanks{%
a.riadsolh@um5r.ac.ma}, M El Belkacemi$^{3}$\thanks{%
mourad\_prof@yahoo.fr} and H. Ez-Zahraouy$^{1,2}$\thanks{%
h.ezzahraouy@um5r.ac.ma} \\
$^{1}${\small Laboratory of Condensend Matter and Interdisplinary Sciences,
Department of physics,}\\
\ {\small Faculty of Sciences, Mohammed V University in Rabat, Morocco}\\
$^{2}${\small CNRST Labeled Research Unit (URL-CNRST), Morocco}\\
$^{3}${\small Laboratory of Conception and Systems (Electronics, Signals and
Informatics)}\\
\ {\small Faculty of Sciences, Mohammed V University in Rabat, Morocco}}
\maketitle

\begin{abstract}
This paper introduces novel solutions for inflation and late-time cosmic
acceleration within the framework of quantum Lovelock gravity, utilizing
Friedmann equations. Furthermore, we demonstrate the hypergeometric states
of cosmic acceleration through the Schr\"{o}dinger stationary equation. A
physical interpretation is proposed, whereby the rescaled Lovelock couplings
represent a topological mass that characterizes the Lovelock branch. This
research holds the potential for an extension into the quantum description.
Predictions for the spectral tilt and tensor-to-scalar ratio are depicted
through plotted curves. By utilizing the rescaled Hubble parameter, the
spectral index is determined in terms of the number of e-folds.
\end{abstract}

\section{Introduction}

Lovelock gravity \cite{Z13} represents a generalization of Einstein's theory
of general relativity to higher dimensions within the realm of higher
curvature gravity theories. Several aspects of Lovelock theories in general
dimensions are studied in \cite{JHEP1,PRD1,PRD2,PRD3,ANN2}, particularly,
the thermodynamic properties of the Lovelock black hole \cite%
{TBH1,TBH2,ANN3,ANN4,ANN5,ANN6}. The articles \cite{PP1,PP2,PP3} focused on
investigating the third-order Lovelock gravity with black holes.
Additionally, there exist other studies examining the connection between
Lovelock gravity and various theories, including Lanczos-Lovelock \cite%
{PO1,PO01}, Lovelock-Cartan theory \cite{PO2}, AdS/CFT \cite{PO3}, and
Horndeski theories \cite{PO4}. In \cite{PO5}, the gravitational field of
Lovelock was investigated within the context of cosmology. The Friedmann
equations are a set of equations that describe the evolution of the universe
on a large scale \cite{FR1,ANL1}. In Lovelock gravity, the Friedmann
equations are modified to include additional terms that arise from the
higher-dimensional nature of the theory \cite{RE1,JHEP2}. Recent progress in
quantum Friedmann equations offers a promising framework for generating new
solutions for cosmic acceleration, as mentioned in \cite{CQG5,PLB3,in3}. The
cosmological Friedman-Einstein dynamical system can be reformulated as a
type of Schr\"{o}dinger equation, where the bound eigensolutions correspond
to oscillating functions \cite{in2}. In quantum gravity, the customary
approach involves treating the entire universe as a quantum system \cite%
{QU0,QU1}. This perspective often entails considering multiple co-existing
and non-interacting universes. The Wheeler-DeWitt equation $\hat{H}\psi =0$,
governing the wave function of the universe $\psi $, can be expressed as a
Schr\"{o}dinger equation within the Friedman-Robertson-Walker (FRW)
spacetime. The article \cite{LF1} discusses an inflationary model that is
based on Lovelock's terms. These terms involve higher-order curvature and
can cause inflation to occur when there are more than three spatial
dimensions.\newline
This study aims to find an interpretation of cosmic expansion following a
connection between Lovelock gravity and quantum mechanics by examining the
energy states of cosmic expansion in terms of Lovelock coupling.\newline
The structure of this paper is outlined as follows: Section 2 delves into
the examination of the Friedmann equations within Lovelock gravity, as well
as the exploration of topological density. In section 3, we study the cosmic
acceleration from quantum states derived by the FRW-Lovelock universe and
Schr\"{o}dinger stationary equation through the Hubble parameter
reparameterization. In section 4, we study the cosmic expansion according to
Schr\"{o}dinger equation and we examine our proposal in various different
hypergeometric states. In Section 5, we analyze the observational
constraints on Lovelock inflation. Finally, section 6 concludes the paper by
summarizing our findings.\ Throughout this article, we use units $c=\hbar
=k_{B}=1$.

\section{General Lovelock gravity}

Let us start with the Lagrangian of the $(D+1)$-dimensional Lovelock gravity
\cite{Z13} of extended Euler densities is given by%
\begin{equation}
\mathcal{L}=\sum_{k=0}^{\bar{D}}\frac{\alpha _{k}}{16\pi G_{N}}2^{-k}\delta
_{c_{1}d_{1}\cdots c_{k}d_{k}}^{a_{1}b_{1}\cdots a_{k}b_{k}}R_{a_{1}b_{1}}^{%
\text{ \ \ \ \ }c_{1}d_{1}}\ldots R_{a_{k}b_{k}}^{\text{ \ \ \ \ }%
c_{k}d_{k}}+\mathcal{L}_{m},  \label{a1}
\end{equation}%
where $L_{m}$\ is the matter Lagrangian, $\bar{D}=\left[ \frac{D-1}{2}\right]
$ (the brackets $\left[ \cdots \right] $ denote taking the integer part), $%
\delta _{c_{1}d_{1}\cdots c_{k}d_{k}}^{a_{1}b_{1}\cdots a_{k}b_{k}}$\ is the
generalized antisymmetric Kronecker delta and $R_{a_{1}b_{1}}^{\text{ \ \ \
\ }c_{1}d_{1}}$ is the Riemann tensor. We notice that $\alpha _{k}$ are the
Lovelock coupling constants with dimensions $(length)^{2k-D}$, where D is
the respective dimension. Here, $\alpha _{2}=\alpha \ $and $\alpha _{3}$ are
respectively the second (Gauss-Bonnet coupling \cite{JCAP4,PLB2}) and the
third Lovelock coefficients. The zero-order Lovelock invariant is $\alpha
_{0}L_{0}=-2\Lambda =\frac{\left( D-1\right) \left( D-2\right) }{L^{2}}$
with $\Lambda $ being the cosmological constant and $L$ is the AdS radius
\cite{JHEAP1}. The first-order Lovelock invariant is $\alpha _{1}L_{1}=R$ is
the Einstein-Hilbert Lagrangian. The equations of motion for Lovelock
gravity \cite{CQG4,PRDQ4} read%
\begin{equation}
\sum_{k=0}^{\bar{D}}\frac{\alpha _{k}}{16\pi G_{N}}2^{-2k-1}\delta _{\mathbf{%
a}a_{1}b_{1}\cdots a_{k}b_{k}}^{\mathbf{b}c_{1}d_{1}\cdots
c_{k}d_{k}}R_{c_{1}d_{1}}^{\text{ \ \ \ \ }a_{1}b_{1}}\ldots =T_{a}^{b}%
\mathbf{,}
\end{equation}%
where $T_{a}^{b}$\ is the stress tensor. Hence, by varying the Euler
density\ \cite{JHEP3}, the stress tensor in a conformally flat background
can be derived as%
\begin{equation}
\left\langle T_{b}^{\text{ \ }a}\right\rangle =-\frac{C_{D}}{\left( -8\pi
\right) ^{\frac{D}{2}}}\lim_{n\rightarrow D}\frac{1}{n-D}R_{\text{ \ \ \ }%
a_{1}a_{2}}^{b_{1}b_{2}}\cdots R_{\text{ \ \ \ }a_{D-1}a_{D}}^{b_{D-1}b_{D}}%
\delta _{b_{1}\cdots b_{D}b}^{a_{1}\cdots a_{D}a},
\end{equation}%
where $C_{D}$ are the central charges. The $\left( D+1\right) $-dimensional
FRW universe with the metric%
\begin{equation}
ds^{2}=-dt^{2}+a^{2}(t)\left( \frac{dr^{2}}{1-\kappa r^{2}}+r^{2}d\Omega
_{D-1}^{2}\right) ,  \label{b1}
\end{equation}%
where $d\Omega _{D-1}^{2}$ denotes the line element of an $(D-1)$%
-dimensional unit sphere. The curvature parameter $\kappa $ can take the
values $\kappa =+1$, $-1$, or $0$, corresponding to closed, open, or flat
universes, respectively. The effective cosmological constant and energy
density are modified in Lovelock gravity due to the presence of
higher-dimensional terms. The pressure $p$ is related to the energy density
and the equation of state of the matter and energy in the universe. In
Lovelock gravity, it is modified by higher-dimensional terms. The continuity
equation of the perfect fluid is given,%
\begin{equation}
\dot{\rho}+HD\left( \rho +p\right) =0,  \label{b2}
\end{equation}%
where $H=\frac{\dot{a}}{a}$ denotes the Hubble parameter and $a\left(
t\right) $ is the scale factor that describes the relative expansion of the
Universe. We introduce the coefficients $c_{k}$ are given by \cite{RE1}%
\begin{eqnarray}
c_{0} &=&\frac{\alpha _{0}}{16\pi G_{N}}\frac{1}{D\left( D-1\right) }\text{,}
\label{b3} \\
c_{1} &=&1,  \notag \\
c_{k} &=&\frac{\alpha _{k}}{16\pi G_{N}}\prod\limits_{i=3}^{2\bar{D}}\left(
D+1-i\right) \text{ for }i>1.  \notag
\end{eqnarray}%
We assume that the entropy of the FRW universe follows the entropy
\begin{equation}
S=\frac{A}{4G_{N}}\sum_{n=1}^{\bar{D}}c_{k}\frac{n\left( D-1\right) }{D-2n+1}%
r_{A}^{2-2n}.  \label{Sen}
\end{equation}%
\ For\ apparent horizon \cite{RE1,PLB1}.Therefore, the radius of the black
hole horizon is substituted with the radius of the apparent horizon \cite%
{PRDQ4} as%
\begin{equation}
r_{A}=\left( H^{2}+\frac{\kappa }{a^{2}}\right) ^{-\frac{1}{2}}.  \label{ra}
\end{equation}%
We consider the entropy of the apparent horizon in the FRW universe,
represented by Eq. (\ref{Sen}). The temperature of the apparent horizon is
written as \cite{CQG00}%
\begin{equation}
T=-\frac{\kappa }{2\pi r_{A}}+\frac{\left( D-1\right) r_{A}}{4\pi \bar{D}%
\left( \frac{\kappa }{r_{A}^{2}}+\left( \bar{D}\alpha \right) ^{-\frac{1}{%
\bar{D}-1}}\right) ^{\bar{D}-1}}\left( \left( \frac{\kappa }{r_{A}^{2}}%
+\left( \bar{D}\alpha \right) ^{-\frac{1}{\bar{D}-1}}\right) ^{\bar{D}%
}-\left( \bar{D}\alpha \right) ^{-\frac{\bar{D}}{\bar{D}-1}}+\frac{\alpha
_{0}}{\alpha }\right) .
\end{equation}%
We consider the standard perfect fluid matter, and apply the first law, $%
dM=TdS$, so that the Klein-Gordon equation can then be written from the
metric function Eq. (\ref{b1}) \cite{RE1} as
\begin{equation}
\sum_{n=1}^{\bar{D}}nc_{n}\left( H^{2}+\frac{\kappa }{a^{2}}\right)
^{n-1}\left( \dot{H}-\frac{\kappa }{a^{2}}\right) =-\frac{8\pi G_{N}}{D-1}%
\left( \rho +p\right) .  \label{b4}
\end{equation}%
Using the continuity Eq. (\ref{b2}) and integrating Eq. (\ref{b4}), we
finally obtain the Friedmann equation in the Lovelock gravity,%
\begin{equation}
\sum_{n=1}^{\bar{D}}c_{n}\left( H^{2}+\frac{\kappa }{a^{2}}\right) ^{n}=%
\frac{16\pi G_{N}}{D\left( D-1\right) }\rho .  \label{b5}
\end{equation}%
The Friedmann equations in Lovelock gravity provide a framework for
understanding the evolution of the universe in higher-dimensional theories
of gravity. These equations can be used to study the behavior of the
universe on large scales and to test the predictions of Lovelock gravity
against observational data. When $\bar{D}=2$, Eqs. (\ref{b4})-(\ref{b5})
reduce to those of the Gauss-Bonnet gravity:%
\begin{eqnarray}
&&\left( 1+2\alpha \left( D-2\right) \left( D-3\right) \left( H^{2}+\frac{%
\kappa }{a^{2}}\right) \right) \left( \dot{H}-\frac{\kappa }{a^{2}}\right)
\label{b6} \\
&=&-\frac{8\pi G_{N}}{\left( D-1\right) }\left( \rho +p\right) ,  \notag
\end{eqnarray}%
\begin{eqnarray}
&&H^{2}+\frac{\kappa }{a^{2}}+\alpha \left( D-2\right) \left( D-3\right)
\left( H^{2}+\frac{\kappa }{a^{2}}\right) ^{2}  \label{b7} \\
&=&\frac{16\pi G_{N}}{D\left( D-1\right) }\rho .  \notag
\end{eqnarray}%
If $\alpha =0$ or $D=3$, we get the standard form of the Friedmann equation.
The Friedmann equations in Lovelock gravity can be used to study the
evolution of the universe during the inflationary period and cosmic
acceleration.

\section{Topological cosmic acceleration}

The effective energy density and pressure of the scalar field are modified
due to the presence of higher-dimensional terms. The scalar field potential
is also modified, which can affect the dynamics of the inflationary period
and the production of primordial gravitational waves. According to the
Einstein-scalar-Gauss-Bonnet gravity, the scalar field can play the role of
Gauss-Bonnet coupling. In the general framework, we can always replace the
Lovelock coupling with a scalar field \cite{JHEP1}. The Brans-Dicke theory
offers a potential explanation for the current accelerated expansion of the
universe, eliminating the need for a cosmological constant or quintessence
matter \cite{PRD4}. The topics of curvature quintessence and cosmic
acceleration can be addressed within frameworks that do not necessarily
involve scalar fields \cite{in1,AX1}. We recall that the deceleration
parameter is $q=-1-\frac{\dot{H}}{H^{2}}$. The accelerated behavior is
achieved if $q<0$. The deceleration parameter can be given in terms of
density $\rho $. First, for comparison, introducing a new reparameterization
in the Lovelock gravity framework: $\sigma =1+\frac{\kappa }{a^{2}H^{2}}$.
The parameter $\sigma $ is described by the curvature parameter $\kappa $.
Using the conformal time $\tau =\sqrt{\sigma }t$, i.e. $\frac{\partial a}{%
a\partial t}=\sqrt{\sigma }\frac{\partial a}{a\partial \tau }$, which can be
used to set $H=\frac{\partial a}{a\partial \tau }$, so the Hubble parameter
simplifies to $H^{2}=\sigma H^{2}$. First, for comparison, introducing a new
reparameterization of Hubble parameter and Lovelock couplings:%
\begin{eqnarray}
\mathcal{H}^{2} &=&\sigma H^{2}=H^{2}+\frac{\kappa }{a^{2}},  \label{bb8} \\
g_{n}^{2} &=&\frac{D\left( D-1\right) }{16\pi G_{N}}c_{n}.  \notag
\end{eqnarray}%
The recent puzzle surrounding the variation in measured values of the Hubble
rate, as observed by the Hubble Space Telescope and the Planck satellite,
adds an intriguing aspect to this topic \cite{Naturea}. This gives the
Hubble parameter great importance for our study, as a free parameter that
generates all the differential equations thereafter. Second, we express the
two Eqs. (\ref{b4},\ref{b5}) in terms of $H$\ as%
\begin{equation}
\rho =\sum_{n=1}^{\bar{D}}g_{n}^{2}\mathcal{H}^{2n}=g_{1}^{2}\mathcal{H}%
^{2}+\cdots +g_{\bar{D}}^{2}\mathcal{H}^{2\bar{D}},  \label{b8}
\end{equation}%
\begin{equation}
\rho +p=-\frac{2\mathcal{\dot{H}}}{DH\mathcal{H}}\sum_{n=1}^{\bar{D}%
}ng_{n}^{2}\mathcal{H}^{2n},  \label{b9}
\end{equation}%
where $\frac{\mathcal{H}}{H}\dot{H}=\dot{H}-\frac{\kappa }{a^{2}}$, $g_{n}$
are rescaled Lovelock coupling and we recall that $H$ is the inverse of
apparent horizon $H^{2}=\frac{1}{r_{A}^{2}}$. The density $\rho $ includes
contributions from all forms of matter and energy in the universe. The
rescaled Hubble parameter checks the symmetry $H\longleftrightarrow -H$ in $%
\rho $. An interesting special case is the similarity between the behavior
of Hubble parameter $H$ in the Friedmann equation and the Higgs field in the
potential, which reveals the Higgs particle mass. According to the Higgs
slow-roll inflation model, the universe underwent a period of exponential
expansion driven by the Higgs field, which acted as a scalar field. During
this period, the Higgs field slowly rolled down its potential energy curve,
releasing energy that drove the inflationary expansion. Next, we assume that
the rescaled Hubble parameter $H$ can describe the cosmic acceleration in
Lovelock gravity. In order to illustrate the similarity between the density $%
\rho $\ and the Higgs potential, let us consider the following potential $%
V_{n}\left( \mathcal{H}\right) =g_{n}^{2}H^{2n}$. Therefore, we can write $%
\rho =V=\sum_{n=1}^{\bar{D}}V_{n}\left( \mathcal{H}\right) $, where $%
V_{n}\left( \mathcal{H}\right) =g_{n}^{2}H^{2n}$. Clearly,%
\begin{eqnarray}
\rho _{0} &=&V_{0}=\frac{\alpha _{0}}{\left( 16\pi G_{N}\right) ^{2}}\text{,
}  \label{b11} \\
\rho _{1} &=&V_{1}=\frac{D\left( D-1\right) }{16\pi G_{N}}\mathcal{H}^{2},
\notag \\
\rho _{q} &=&V_{q}=\frac{\alpha _{q}D\left( D-1\right) }{\left( 16\pi
G_{N}\right) ^{2}}\mathcal{H}^{2q}\prod\limits_{i=3}^{2\bar{D}}\left(
D+1-i\right) \text{ for }i>1,  \notag
\end{eqnarray}%
where $1<i$ and $1<q<\bar{D}$. If we take for example $\left\vert \mathcal{H}%
\right\vert ^{2}=H^{2}+\frac{\left\vert \kappa \right\vert }{a^{2}}$, then $%
\left\vert \kappa \right\vert =\delta =\left( 0,1\right) $, in this case,
the rescaled Hubble parameter $H$ becomes complex (ex: superfield) which is
useful for inflation in supergravity \cite{PRL2}. Note that $H$ and $H^{\ast
}$ should be taken as $H=H+i\frac{\delta }{a^{2}}$ and $H^{\ast }=H-i\frac{%
\delta }{a^{2}}$. On the other hand, the potentials $\left\{
V_{0},V_{1},\cdots ,V_{\bar{D}}\right\} $ encompass a range of cosmic
expansion spectra; however, to obtain a comprehensive understanding,
additional information is required. As we begin with an energy spectrum, it
is necessary to incorporate quantum states, assigning unique energies to
each state, in order to evaluate this spectrum from a quantum perspective.
We introduce the effective thermodynamic pressure \cite{CQG2} as $p=\frac{%
\alpha _{0}}{16\pi G_{N}}=-\frac{\Lambda }{8\pi G_{N}}$. This implies that
the density $\rho _{0}$ represents the pressure in space-time: $p=16\pi
G_{N}\rho _{0}$. In this case, we can write the equations of state as%
\begin{eqnarray}
\omega _{0} &=&\frac{p}{\rho _{0}}=16\pi G_{N}\text{, \ \ }\omega _{1}=\frac{%
\alpha _{0}}{D\left( D-1\right) \mathcal{H}^{2}},  \label{b12} \\
\omega _{q} &=&\frac{16\pi G_{N}\alpha _{0}}{\alpha _{q}D\left( D-1\right)
\mathcal{H}^{2q}\prod\limits_{i=3}^{2\bar{D}}\left( D+1-i\right) }.  \notag
\end{eqnarray}%
Note that the equations of state $\omega _{q\geq 1}$ depend on the Hubble
parameter and the Lovelock coupling. It is important to note that the
potential $V_{0}$ in this context corresponds to the pressure $p$: $V_{0}=%
\frac{1}{\alpha _{0}}p^{2}$. Since the potential $V_{1}$ is proportional to
the square of the Hubble parameter, this indicates that $V_{1}$ is a
density. Having deduced these potentials, so we interpret $\left\{
V_{0},V_{1},\cdots ,V_{\bar{D}}\right\} $ as a potential spectrum that
describes the cosmic expansion. We can rewrite Eqs. (\ref{b11}) to 4
dimensions as follows:%
\begin{eqnarray}
\rho _{0} &=&\frac{\alpha _{0}}{\left( 16\pi G_{N}\right) ^{2}}\text{, }\rho
_{1}=\frac{12\mathcal{H}^{2}}{16\pi G_{N}},\text{ }\rho _{2}=\frac{24\alpha
_{2}\mathcal{H}^{4}}{\left( 16\pi G_{N}\right) ^{2}},  \label{b13} \\
\omega _{0} &=&16\pi G_{N}\text{, \ }\omega _{1}=\frac{\alpha _{0}}{12%
\mathcal{H}^{2}},\text{ \ }\omega _{2}=\frac{2\pi G_{N}\alpha _{0}}{3\alpha
_{2}\mathcal{H}^{4}}.  \notag
\end{eqnarray}%
In the limit of $\alpha _{k\geq 2}\rightarrow 0$, Lovelock gravity reduces
to Einstein gravity \cite{REV1}, implying that all the densities become zero
except for two densities $\rho _{0}$\ and $\rho _{1}$. The density $\rho
_{0} $\ arises from the cosmological constant, while the density $\rho _{1}$%
\ corresponds to the Friedman equations. The static nature of the universe
is described by the density $\rho _{0}$, while the density $\rho _{1}$
signifies the geometric expansion acceleration indicated by $H^{2}$ (Hubble
parameter and the curvature parameter). In contrast, the
geometric/topological density $\rho _{2}$ represents an acceleration of
expansion resulting from the term $H^{4}$ and the topological parameter $%
\alpha =\alpha _{2}$. These results indicate that the equations of state $%
\omega _{0}$ are not influenced by the topology of space-time. Conversely,
the equations of state $\omega _{1}$ and $\omega _{2}$ explicitly depend on $%
\alpha _{0}\sim -\Lambda $, which describes the cosmological constant. While
$\omega _{2}$ decreases with $\alpha _{2}$. The presence of two types of
expansion acceleration: $\rho _{2}\sim H^{4}$ and $\rho _{1}\sim H^{2}$ in
Eqs. (\ref{b13}) is in good agreement with inflation expansion and late-time
cosmic acceleration. Therefore, Lovelock's model can unify these two
phenomena.

\section{Hypergeometric\ states}

\begin{figure}[H]
\centering\includegraphics[width=11cm]{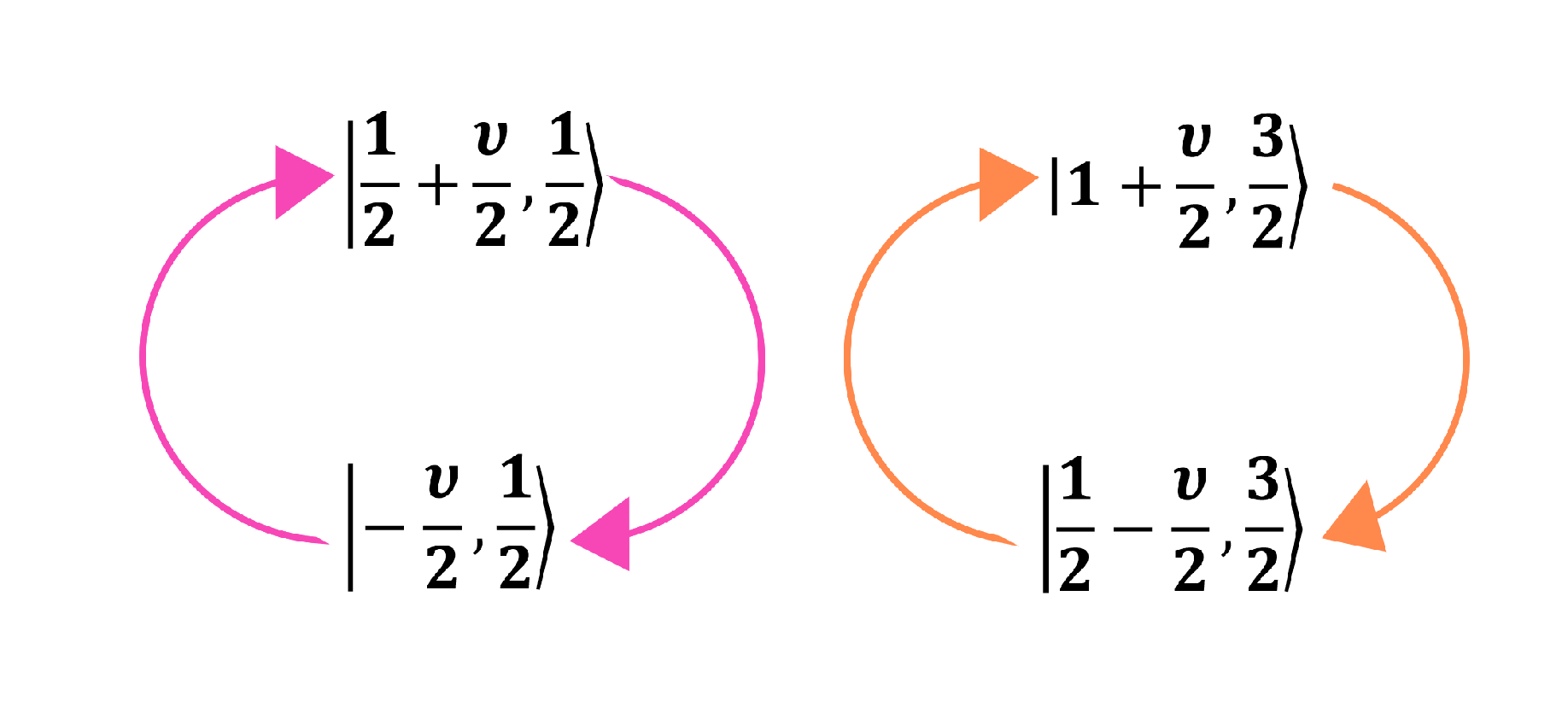}
\caption{Loop of repeating states $\left\vert -\frac{\protect\upsilon }{2},%
\frac{1}{2}\right\rangle \circlearrowleft \left\vert \frac{1}{2}+\frac{%
\protect\upsilon }{2},\frac{1}{2}\right\rangle $ and $\left\vert \frac{1}{2}-%
\frac{\protect\upsilon }{2},\frac{3}{2}\right\rangle \circlearrowleft
\left\vert 1+\frac{\protect\upsilon }{2},\frac{3}{2}\right\rangle $.}
\label{F1}
\end{figure}

Within the framework of cosmic acceleration, we introduce the wave function $%
\Psi =\Psi \left( a,t\right) $ that assigns the cosmological scale factor $%
a(t)$ at each time $t$. Consequently, we derive the Schr\"{o}dinger equation
\cite{in2,in4,in5}:%
\begin{equation}
i\frac{\partial \Psi }{\partial t}=-\frac{1}{2m}\frac{\partial ^{2}\Psi }{%
\partial a^{2}}+V(a)\Psi .  \label{sc1}
\end{equation}%
Using $a/a_{0}=\left( 1+z\right) ^{-1}$ then, $\Psi =\Psi \left( a,t\right)
=\Psi \left( z,t\right) $ is the formal probability amplitude to find a
given galaxy of mass $m$ at a given $a(t)$ or a given redshift $z$ at time $t
$ \cite{in2}. In our specific scenario, the mass $m=4\pi r^{3}\rho /3$ can
be interpreted as the mass of a galaxy\ \cite{in2,in3}, while the quantity $%
\left\vert \Psi \right\vert ^{2}$ represents the probability of encountering
such a galaxy. We consider a stationary state with energy $E$ as $\Psi
\left( a,t\right) =\psi \left( a\right) e^{-iEt}$, and the Schr\"{o}dinger
stationary equation is%
\begin{equation}
\frac{\partial ^{2}\psi \left( H\right) }{\partial H^{2}}+\frac{2m}{H^{2}}%
\left( E-V\right) \psi \left( H\right) =0,  \label{sc2}
\end{equation}%
where $\left( \partial H/\partial a\right) ^{2}=H^{2}$, i.e. $\frac{\partial
^{2}\psi \left( H\right) }{\partial a^{2}}=H^{2}\frac{\partial ^{2}\psi
\left( H\right) }{\partial H^{2}}$. If we assume that $m$ does not depend on
$H$, the normalized ground state wave function can be written as%
\begin{equation}
\psi \left( H\right) =\psi _{0-}H^{\frac{1}{2}\left( 1-\Delta \right) }+\psi
_{0+}H^{\frac{1}{2}\left( 1+\Delta \right) },  \label{sc3}
\end{equation}%
where $\Delta =\sqrt{1-8m\left( E-V\right) }$, i.e. $E=\frac{1}{8\rho
\mathcal{V}}\left( 1-\Delta ^{2}\right) +\rho $ and $V=\frac{4\pi r^{3}}{3}$%
. This solution can be expressed as the combination of the function $\psi
_{-}=\psi _{0-}H^{\frac{1}{2}\left( 1-\Delta \right) }$ that travels to the
right and the function $\psi _{+}=\psi _{0+}H^{\frac{1}{2}\left( 1+\Delta
\right) }$ that travels to the left. The solution represented by Eq. (\ref%
{sc3}) does not depict the standing wave within the framework of Lovelock
gravity. In order to express the solution in line with Lovelock gravity, we
need to incorporate the relationship between mass and density: $m=4\pi
r^{3}\rho /3$. From Eqs. (\ref{bb8},\ref{b8},\ref{sc2}):%
\begin{equation}
\frac{\partial ^{2}\psi \left( H\right) }{\partial H^{2}}+\frac{2\mathcal{V}%
}{H^{2}}\left( E-V\right) \sum_{n=1}^{\bar{D}}g_{n}^{2}\mathcal{H}^{2n}\psi
\left( H\right) =0.  \label{sc5}
\end{equation}%
In the regime of $E=V$, we find $\psi \left( H\right) =aH+b$. The conditions
$a=1$ and $b=0$ implies that $\psi \left( H\right) =H$. If we are in some
regime where $V\ll E$, $\sigma =1$ ( or $\kappa =0$) and $D=4$, Eq. (\ref%
{sc5}) can be written as%
\begin{equation}
\frac{\partial ^{2}\psi \left( H\right) }{\partial H^{2}}+2\mathcal{V}%
E\left( g_{1}^{2}+g_{2}^{2}H^{2}\right) \psi \left( H\right) =0.  \label{sc6}
\end{equation}%
The general solution can be written as (see Appendix A)
\begin{equation}
\psi \left( H\right) =\psi _{0-}\mathcal{D}_{\bar{\upsilon}}\left( \left(
1+i\right) HB\right) +\psi _{0+}\mathcal{D}_{\upsilon }\left( -\left(
1-i\right) HB\right) .  \label{sc7}
\end{equation}%
Here, $\mathcal{D}$ is the parabolic cylinder function \cite{JCAM,PLA},
where $\upsilon =-\frac{1}{2}+\frac{ig_{1}^{2}}{2}\sqrt{\frac{2E\mathcal{V}}{%
g_{2}^{2}}}$, $\bar{\upsilon}=-\frac{1}{2}-\frac{ig_{1}^{2}}{2}\sqrt{\frac{2E%
\mathcal{V}}{g_{2}^{2}}}$ and $B=\left( 2E\mathcal{V}g_{2}^{2}\right) ^{%
\frac{1}{4}}$. Note that we have found the general solution in terms of the
parameter $\upsilon $. Consequently, we can easily go from the total energy
to the parameter $\upsilon $ using the following relation:%
\begin{equation}
E=\frac{\alpha _{2}}{3\mathcal{V}}\left( \left\vert \upsilon \right\vert
^{2}-\frac{1}{4}\right) ,  \label{sc8}
\end{equation}%
where $\frac{g_{2}^{2}}{g_{1}^{4}}=\frac{\alpha _{2}}{6}$. Note that the
total energy of the$\ \left( 4+1\right) $-dimensional system depends on the
Gauss-Bonnet coupling $\alpha _{2}=\alpha $. This relationship suggests that
the geometry of the\ $\left( 4+1\right) $-dimensional universe, as
determined by the Gauss-Bonnet term, has an impact on the dynamics of cosmic
expansion. This implies that the energy of cosmic expansion depends on the
space-time topology Eqs. (\ref{b13}). The zero state possesses no energy and
is distinguished by $\upsilon =\frac{1}{2}$. The parabolic cylinder function
can be expressed in terms of confluent hypergeometric functions $\mathcal{M}$
as \cite{MAT1,MAT2,MAT3,MAT4}%
\begin{eqnarray}
\mathcal{D}_{\upsilon }\left( z\right)  &=&\frac{1}{\sqrt{\pi }}2^{\frac{%
\upsilon }{2}}e^{-\frac{z^{2}}{4}}\times \left( \cos \left( \frac{\pi
\upsilon }{2}\right) \right.   \label{sc9} \\
&&\times \Gamma \left( \frac{1+\upsilon }{2}\right) \mathcal{M}\left( -\frac{%
\upsilon }{2};\frac{1}{2};\frac{z^{2}}{2}\right)  \\
&&+\sqrt{2}z\sin \left( \frac{\pi \upsilon }{2}\right)   \notag \\
&&\times \left. \Gamma \left( \frac{2+\upsilon }{2}\right) \mathcal{M}\left(
\frac{1}{2}-\frac{\upsilon }{2};\frac{3}{2};\frac{z^{2}}{2}\right) \right) .
\end{eqnarray}%
We note that the variable $z$ depends on the Hubble parameter as $z=\left(
-1+i\right) HB$. The parabolic cylinder function describes the behavior of
waves or wave-like phenomena that exhibit cylindrical symmetry and possess
parabolic characteristics with cylindrical symmetry. The parabolic cylinder
function provides a mathematical representation of the space-time
distribution of a hypergeometric state. It determines the shape of the
wavefronts and how the energy of the waves is distributed within the
hypergeometric state. The analysis of asymptotic behavior results is
particularly interesting. First, we distinguish between two fundamental
cases: $\upsilon =\left\{ 1,2\right\} $:%
\begin{eqnarray}
\mathcal{D}_{1}\left( z\right)  &=&+ze^{-\frac{z^{2}}{4}}\mathcal{M}\left( 0;%
\frac{1}{2}+1;\frac{z^{2}}{2}\right) ,  \label{sc10} \\
\mathcal{D}_{2}\left( z\right)  &=&-e^{-\frac{z^{2}}{4}}\mathcal{M}\left( -1;%
\frac{1}{2};\frac{z^{2}}{2}\right) ,  \label{sc11}
\end{eqnarray}%
where $\Gamma \left( \frac{3}{2}\right) =\sqrt{\pi }/2$. Here, $\mathcal{M}%
\left( a;b;z\right) =\sum_{n=0}^{\infty }\frac{a^{\left( n\right) }z^{n}}{%
b^{\left( n\right) }n!}$ with $a^{\left( 0\right) }=1$ and $a^{\left(
n\right) }=a\left( a+1\right) \left( a+2\right) \cdots \left( a+n-1\right) $%
, is the rising factorial. Clearly, $\mathcal{D}_{1}\left( z\right) \sim
\partial _{z}\mathcal{D}_{2}\left( z\right) $ and to pass from $\mathcal{M}$
of $\mathcal{D}_{2}$ to $\mathcal{M}$ of $\mathcal{D}_{1}$, it is enough to
add to $a$ and $b$ the number $1$, but not for $\frac{z^{2}}{2}$. So we
interpret $\mathcal{M}$-functions as creation and annihilation operators.
Using Kummer's transformations \cite{CQG6}: $\mathcal{M}\left( a,b,z\right)
=e^{z}\mathcal{M}\left( b-a,b,-z\right) $, we can generate a series of
functions $\mathcal{M}$ present in Eq. (\ref{sc9}). We first choose the
state $\mathcal{M}\left( -\frac{\upsilon }{2};\frac{1}{2};\frac{z^{2}}{2}%
\right) $,%
\begin{eqnarray}
&&\vdots   \notag \\
\mathcal{M}\left( -\frac{\upsilon }{2};\frac{1}{2};\frac{z^{2}}{2}\right)
&=&e^{+\frac{z^{2}}{2}}\mathcal{M}\left( \frac{1}{2}+\frac{\upsilon }{2};%
\frac{1}{2};-\frac{z^{2}}{2}\right) , \\
\mathcal{M}\left( \frac{1}{2}+\frac{\upsilon }{2};\frac{1}{2};\frac{z^{2}}{2}%
\right)  &=&e^{+\frac{z^{2}}{2}}\mathcal{M}\left( -\frac{\upsilon }{2};\frac{%
1}{2};-\frac{z^{2}}{2}\right) .
\end{eqnarray}%
We notice that according to these series, we can create looped states $%
\left\vert -\frac{\upsilon }{2}\right\rangle \rightarrow \left\vert \frac{1}{%
2}+\frac{\upsilon }{2}\right\rangle \rightarrow \left\vert -\frac{\upsilon }{%
2}\right\rangle $. The loop $\left\vert -\frac{\upsilon }{2},\frac{1}{2}%
\right\rangle \circlearrowleft \left\vert \frac{1}{2}+\frac{\upsilon }{2},%
\frac{1}{2}\right\rangle $ repeats infinitely Fig. \ref{F1}. Secondly, we
take $\mathcal{M}\left( \frac{1}{2}-\frac{\upsilon }{2};\frac{3}{2};\frac{%
z^{2}}{2}\right) $ from Eq. (\ref{sc9}):%
\begin{eqnarray}
&&\vdots   \notag \\
\mathcal{M}\left( \frac{1}{2}-\frac{\upsilon }{2};\frac{3}{2};\frac{z^{2}}{2}%
\right)  &=&e^{+\frac{z^{2}}{2}}\mathcal{M}\left( 1+\frac{\upsilon }{2};%
\frac{3}{2};-\frac{z^{2}}{2}\right) , \\
\mathcal{M}\left( 1+\frac{\upsilon }{2};\frac{3}{2};\frac{z^{2}}{2}\right)
&=&e^{+\frac{z^{2}}{2}}\mathcal{M}\left( \frac{1}{2}-\frac{\upsilon }{2};%
\frac{3}{2};-\frac{z^{2}}{2}\right) .
\end{eqnarray}%
Alternatively, this series suggests the presence of a recurring state loop. $%
\left\vert \frac{1}{2}-\frac{\upsilon }{2},\frac{3}{2}\right\rangle
\circlearrowleft \left\vert 1+\frac{\upsilon }{2},\frac{3}{2}\right\rangle $%
. The existence of three parameters in the state function $\mathcal{M}$\
suggests the existence of three quantum numbers $(l,m,n)$\ of $\mathcal{M}%
\left( l,m,n\right) $\ that describe cosmic acceleration:%
\begin{eqnarray}
l &=&\left\{ -\frac{\upsilon }{2},\frac{1}{2}-\frac{\upsilon }{2},\frac{1}{2}%
+\frac{\upsilon }{2},1+\frac{\upsilon }{2}\right\} , \\
m &=&\left\{ \frac{1}{2},\frac{3}{2}\right\} , \\
n &\mathbf{=}&\left\{ \mathbf{-}\frac{z^{2}}{2},+\frac{z^{2}}{2}\right\} .
\end{eqnarray}%
For $\left( \upsilon =\frac{1}{2},z=\pm 1\right) $\ we get $l=\left\{ -\frac{%
1}{4},\frac{1}{4},\frac{3}{4},\frac{5}{4}\right\} $, $m=\left\{ \frac{1}{2},%
\frac{3}{2}\right\} $\ and $n=\left\{ \mathbf{-}\frac{1}{2},+\frac{1}{2}%
\right\} $. If $\left( \upsilon =1,z=\pm 1\right) $\ we obtain $l=\left\{ -%
\frac{1}{2},0,1,\frac{3}{2}\right\} ,m=\left\{ \frac{1}{2},\frac{3}{2}%
\right\} ,n=\left\{ \mathbf{-}\frac{1}{2},+\frac{1}{2}\right\} $. Note that
the number n does not depend on the parameters $(\upsilon ,z)$. Furthermore,
if we take the loop $\left\vert -\frac{\upsilon }{2}\right\rangle
\rightarrow \left\vert \frac{1}{2}+\frac{\upsilon }{2}\right\rangle
\rightarrow \left\vert -\frac{\upsilon }{2}\right\rangle $, the cyclic
nature of these states implies that the phenomenon of cosmic acceleration,
observed during the inflationary period, has reoccurred in the form of
late-time acceleration. This observation is indicated by the negative sign $%
(-)$\ in the state $\left\vert -\frac{\upsilon }{2}\right\rangle $, which
characterizes the state equation of dark energy (ex: $\omega =-1$).
Moreover, the state $\left\vert \frac{1}{2}+\frac{\upsilon }{2}\right\rangle
$\ refers to the radiation era and the matter era, as the positive value of
the quantum number $\frac{1}{2}+\frac{\upsilon }{2}$\ is directly linked to $%
\omega _{rad}=1/3$\ and $\omega _{matter}=0$.
The parameter $\upsilon $ depend on the energy of the cosmic states and the
space-time topology Eq. (\ref{sc8}). Whereas the parameter $z$ depends on
the Hubble parameter. The two parameters ($\upsilon ,z$) are complex
numbers, but they have an importance in our study. These parameters possess
the remarkable ability to generate comprehensive information about the
cosmic states.

\section{ Constraining Lovelock inflation by observation}

Let us now briefly address the physical properties of the Lovelock inflation
\cite{LF1}. We note that cosmic inflation is a theory that describes the
early universe in which the universe underwent a period of extremely rapid
expansion \cite{JCAP1,JCAP2,JCAP3}. In Lovelock gravity, the theory of
inflation is modified due to the presence of higher-dimensional terms Eq. (%
\ref{b8}). To make a possible generation of new solutions for inflation in
Lovelock gravity with the rescaled Lovelock coupling $H$,\ we define the
slow roll parameter $\epsilon _{\mathcal{H}}=-\frac{\mathcal{\dot{H}}}{%
\mathcal{H}^{2}}\approx -\frac{\dot{H}-\frac{\kappa }{a^{2}}}{H^{2}+\frac{%
\kappa }{a^{2}}}$. The $\epsilon _{H}$ parameter is approximately given by $%
\epsilon _{\mathcal{H}}\approx \epsilon _{H}=-\dot{H}/H^{2}\ll 1$. Leads to%
\begin{equation}
\mathcal{\dot{H}}=-\epsilon _{H}H\mathcal{H}.  \label{c6}
\end{equation}%
Next, we calculate the derivative of $\epsilon _{H}$ by using Eq. (\ref{c6})$%
\ \dot{\epsilon}_{H}=\dot{H}\epsilon _{H}^{2}-\frac{\mathcal{\ddot{H}}}{H%
\mathcal{H}}$. Following \cite{L15}, the slow-roll\ parameter is given by $%
\eta =\epsilon _{H}-\frac{\dot{\epsilon}_{H}}{2H\epsilon _{H}}$. In the
conventional inflationary scenario, the $\epsilon _{H}$ grows as well $%
d\epsilon _{H}/dN=2\epsilon _{H}\left( \eta -\epsilon _{H}\right) $. In this
case, the spectral index of perturbations is equal to $n_{s}=1-6\epsilon
_{H}+2\eta $ \cite{L15,PRD2}. We can estimate the spectral index by the
ratio: $n_{s}=1-4\epsilon _{H}-\frac{\dot{H}}{H}\epsilon _{H}+\frac{\mathcal{%
\ddot{H}}}{\mathcal{H}H^{2}\epsilon _{H}}.$ Define the first potential
slow-roll parameter $\epsilon _{V}=\frac{1}{16\pi G_{N}}\left[ \frac{%
V^{\prime }}{V}\right] ^{2}.$ One obtains $\epsilon _{V}=\frac{1}{16\pi G_{N}%
}\frac{n^{2}}{\mathcal{H}^{2}}$ where $\dot{V}=\frac{\partial V}{\partial t}$
and $V^{\prime }=\frac{\partial V}{\partial \mathcal{H}}$. We thus introduce
the tensor-to-scalar ratio%
\begin{equation}
r=16\epsilon _{V}=\frac{n^{2}}{\pi G_{N}\mathcal{H}^{2}}.  \label{c9}
\end{equation}%
The number of e-foldings counted backward from the end of inflation can be
estimated as $N=\int_{\mathcal{H}}^{\mathcal{H}_{end}}\frac{H}{\mathcal{\dot{%
H}}}dH$ \cite{PRD2}. The number $N$ describes the amount of expansion that
occurs during the inflationary period and is related to the ratio of the
final and initial sizes of the universe. The field value $H_{f}$ at the end
of inflation is determined by the condition $\epsilon _{H}=1$. Using $\dot{H}%
=\epsilon _{H}HH$. Then, the number of e-folds can be given by
\begin{equation}
N=\frac{1}{\epsilon _{H}}\int_{\mathcal{H}}^{\mathcal{H}_{end}}d\ln \mathcal{%
H}=\frac{1}{\epsilon _{H}}\ln \frac{\mathcal{H}_{end}}{\mathcal{H}}.
\label{cn}
\end{equation}%
The effective energy density and pressure of the scalar field are modified,
which can affect the dynamics of inflation and the value of the Hubble
parameter during the inflationary period. This, in turn, can affect the
value of $N$. On using Eq. (\ref{cn}), we can express the rescaled Lovelock
coupling in the forms%
\begin{equation}
\mathcal{H}=\mathcal{H}_{end}e^{-N\epsilon _{H}}.  \label{c10}
\end{equation}%
From \cite{CQG3}, the term $e^{-2N}=\left( \frac{\mathcal{H}}{\mathcal{H}%
_{end}}\right) ^{\frac{2}{\epsilon _{H}}}$ describes the ratio between the
initial and the final phase of inflation. The evolution of the rescaled
Lovelock coupling is governed by $\frac{\mathcal{\dot{H}}}{\mathcal{H}}%
=-\left( \dot{N}\epsilon _{H}+N\dot{\epsilon}_{H}\right) $. It is worth
noticing that $r=\frac{n^{2}}{\pi G_{N}\mathcal{H}_{end}^{2}}e^{2N\epsilon
_{H}}$. The e-folds before the end of inflation for the formation of the
visible part of the Universe is $N=60$ \cite{L12,L13,L14}, i.e. $\dot{N}=0$.
From this and Eq. (\ref{c6}) we find $N\dot{\epsilon}_{H}=H\epsilon _{H}$.
Thus, we can find a useful relation between $\eta $ and $N$ as $\eta
=\epsilon _{H}-\frac{1}{2N}$. The spectral index reduces to%
\begin{equation}
n_{s}=1-4\epsilon _{H}-\frac{1}{N}.  \label{c13}
\end{equation}%
This expression is in good agreement with that found in \cite{JCAP2}. The
Planck CMB data implies that $n_{s}\approx 0.965\pm 0.0042$ \cite{L2}. For
simplicity, Eq. (\ref{c10}) can be approximately obtained as $1-\frac{%
\mathcal{H}}{\mathcal{H}_{end}}\approx N\epsilon _{H}$, or equivalently $%
\left( 1-n_{s}\right) N=1+4\left( 1-\frac{\mathcal{H}}{\mathcal{H}_{end}}%
\right) $. We find%
\begin{equation}
n_{s}=1-\frac{5}{N}+\frac{4\mathcal{H}}{N\mathcal{H}_{end}}.  \label{c14}
\end{equation}%
We first focus on $N=60$. In particular, for the case with $H=H_{end}$, we
have $n_{s}\left( \mathcal{H}_{end}\right) =1-\frac{1}{N}\approx 0.983$, and
for the case with $H=2H_{end}$, we have $n_{s}\left( 2\mathcal{H}%
_{end}\right) \approx 1+\frac{3}{N}=1.05$. Additionally, for $H_{\min }=%
\frac{1}{2}H_{end}$ and $H_{\min }=\frac{3}{4}H_{end}$ we have $n_{s}=0.95$
and $n_{s}=0.96$, respectively. As a result, the observational constraint
derived from the recent Planck 2018 data \cite{L2} already rules out the
class of the inflation model when $\frac{1}{2}H_{end}<H<H_{end}$. From this
condition we get $\frac{1}{2}<e^{-N\epsilon _{H}}<1$, which leads to $%
0<\epsilon _{H}<\frac{1}{2N}$.
\begin{figure}[H]
\centering\includegraphics[width=13cm]{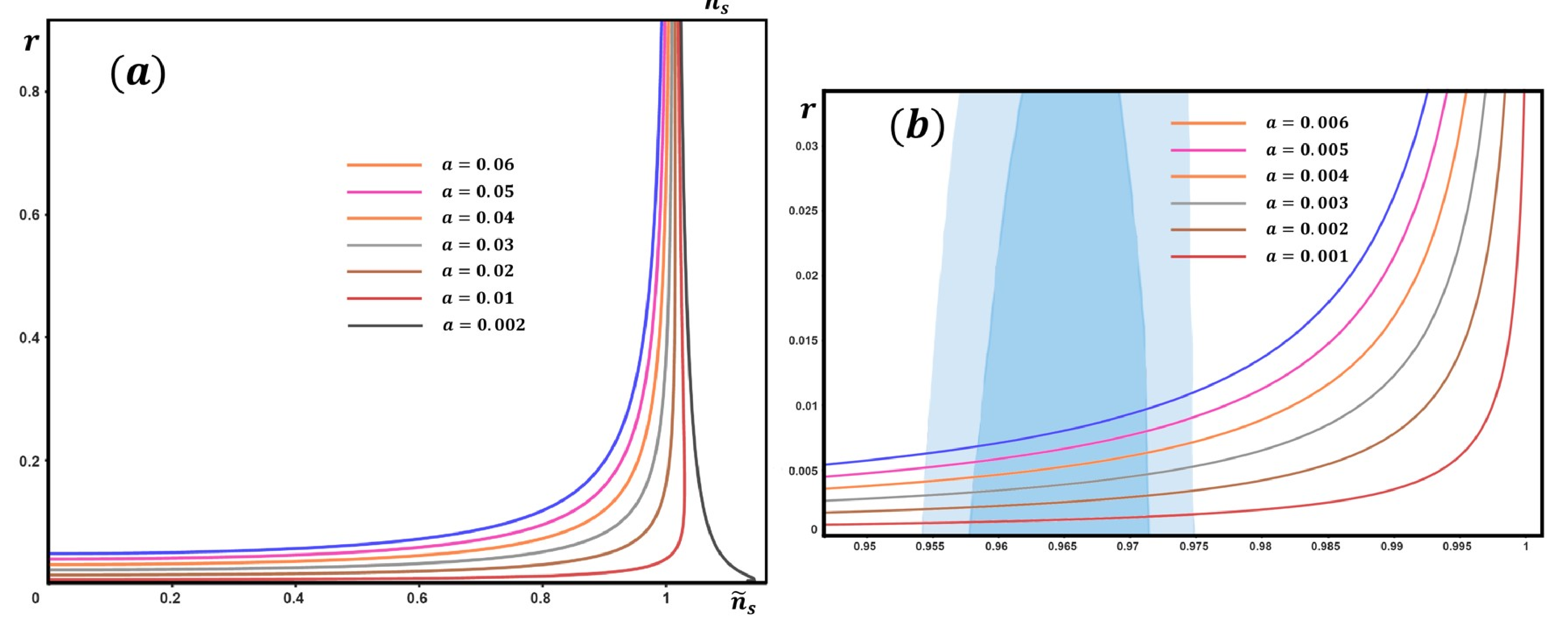}
\caption{(a) The spectral index $\tilde{n}_{s}$ as a function of $r$ for $%
b=0.05$. (b) The curves correspond to the spectral index $\tilde{n}_{s}$ as
a function of $r$ for $b=0.0005$. The Lovelock gravity framework yields
predictions for the spectral tilt and tensor-to-scalar ratio within the
context of inflation.}
\label{F2}
\end{figure}
One can further define a new spectral index $\tilde{n}_{s}=n_{s}+\frac{5}{N}%
=n_{s}+0.083$, such that Eq. (\ref{c14}) can be expressed $\tilde{n}_{s}-%
\frac{4}{N\mathcal{H}_{end}}H=1.$ Let us comment on the spectral index in
the monomial inflation. From Eqs. (\ref{c9},\ref{c14}) we have $\tilde{n}%
_{s}^{2}+\frac{16}{N^{2}\mathcal{H}_{end}^{2}}H^{2}-\frac{8\tilde{n}_{s}%
\mathcal{H}}{N\mathcal{H}_{end}}=1,$or equivalently,%
\begin{equation}
\tilde{n}_{s}^{2}+\frac{16n^{2}}{\pi G_{N}N^{2}\mathcal{H}_{end}^{2}}\frac{1%
}{r}-\frac{8\tilde{n}_{s}n}{N\mathcal{H}_{end}}\sqrt{\frac{1}{\pi G_{N}r}}=1.
\label{c16}
\end{equation}%
From which one finds that $\tilde{n}_{s}^{2}+\frac{a}{r}-\frac{b}{\sqrt{r}}%
=1 $, where $a=\frac{16n^{2}}{\pi G_{N}N^{2}\mathcal{H}_{end}^{2}}$ and $b=%
\frac{8\tilde{n}_{s}n}{N\mathcal{H}_{end}\sqrt{\pi G_{N}}}$. From which, we
plot the predictions in the $\tilde{n}_{s}$ vs $r$ plane for the case $%
b=0.05 $ see Fig. \ref{F2}-(a) and $b=0.0005$ in Fig. \ref{F2}-(b).

\section{Conclusion}

To summarize, this study examines the properties of cosmic acceleration
within the Friedmann-Robertson-Walker metric, considering the density,
potential, and equation of state in relation to the Hubble parameter and the
Lovelock coupling. The energy states of cosmic acceleration are explored
through the Lovelock densities and the Schr\"{o}dinger equation. The
rescaled Hubble parameter directly linked to cosmic acceleration by the
energy density, is shown to be crucial for understanding the dynamics of the
universe. Quantum states are found to follow a parabolic cylinder function,
with hypergeometric functions acting as creation and annihilation operators.
A family of symmetric numbers is proposed to generate the energy spectrum of
cosmic acceleration. We have found that there is a possibility of looping
the quantum states of cosmic acceleration. The analysis focuses on specific
hypergeometric states in Lovelock gravity, revealing important insights into
the kinetic energy and mass within the density parameters. Finally, the
study discusses the diverse quantum properties associated with cosmic
acceleration.\newline
By exploring the predicted spectral tilt and tensor-to-scalar ratio, the
study offers valuable insights into the observational consequences of the
proposed models. The plotted curves effectively depict these predictions,
providing a visual representation of the theoretical findings. An essential
aspect of this work is the determination of the spectral index in terms of
the number of e-folds, achieved through the utilization of the obtained
inflaton field. The proposed concept of Lovelock couplings as representing a
topological mass, which characterizes the Lovelock branch, introduces a
novel perspective on the underlying physics and offers potential for further
exploration, including a quantum description.\newline
In summary, this research provides valuable insights into chaotic inflation
within Lovelock gravity, with implications for our understanding of the
early universe and its inflationary dynamics.\newline
Acknowledgments\newline
The authors thanks Andronikos Paliathanasis for useful enlightening
discussions on the subject of cosmic acceleration and inflaton field. We
thank Neda Farhangkhah and Liu Zhao for their detailed comments on the
Lovelock gravity and dark energy.

\section{Appendix A}

From Eq. (\ref{sc2}) we have%
\begin{equation}
\frac{\partial ^{2}\psi \left( H\right) }{\partial H^{2}}+\frac{2m}{H^{2}}%
\left( E-V\right) \psi \left( H\right) =0.  \label{App1}
\end{equation}%
Using $m=\rho V$\ we obtain
\begin{equation}
\frac{\partial ^{2}\psi \left( H\right) }{\partial H^{2}}+\frac{2\rho
\mathcal{V}}{H^{2}}\left( E-V\right) \psi \left( H\right) =0.  \label{App2}
\end{equation}%
Substituting Eq. (\ref{b8}): $\rho =\sum_{n=1}^{\bar{D}}g_{n}^{2}H^{2n}$\ in
Eq. (\ref{App2}) we get%
\begin{equation}
\frac{\partial ^{2}\psi \left( H\right) }{\partial H^{2}}+\frac{2\mathcal{V}%
}{H^{2}}\left( E-V\right) \sum_{n=1}^{\bar{D}}g_{n}^{2}\mathcal{H}^{2n}\psi
\left( H\right) =0,  \label{App3}
\end{equation}%
or equivalently%
\begin{equation}
\frac{\partial ^{2}\psi \left( H\right) }{\partial H^{2}}+\frac{2\mathcal{V}%
}{H^{2}}\left( E-V\right) \sum_{n=1}^{\bar{D}}\left( g_{1}^{2}\mathcal{H}%
^{2}+\cdots +g_{\bar{D}}^{2}\mathcal{H}^{2\bar{D}}\right) \psi \left(
H\right) =0.  \label{App4}
\end{equation}%
On the other hand from Eq. (\ref{bb8}): $H^{2}=H^{2}+\frac{\kappa }{a^{2}}$%
and Eq. (\ref{App3}) we obtain%
\begin{equation}
\frac{\partial ^{2}\psi \left( H\right) }{\partial H^{2}}+\frac{2\mathcal{V}%
}{H^{2}}\left( E-V\right) \sum_{n=1}^{\bar{D}}g_{n}^{2}\left( H^{2}+\frac{%
\kappa }{a^{2}}\right) ^{n}\psi \left( H\right) =0.  \label{App5}
\end{equation}%
In the regime of $V\ll E$, $\kappa =0$\ and $D=4$ , i.e. $\bar{D}=2$, Eq. (%
\ref{App5}) can be written as%
\begin{equation}
\frac{\partial ^{2}\psi \left( H\right) }{\partial H^{2}}+2\mathcal{V}%
E\left( g_{1}^{2}+g_{2}^{2}H^{2}\right) \psi \left( H\right) =0.
\label{App6}
\end{equation}%
The solution of Eq. (\ref{App6}) can be written as
\begin{equation}
\psi \left( H\right) =\psi _{0-}\mathcal{D}_{\bar{\upsilon}}\left( \left(
1+i\right) HB\right) +\psi _{0+}\mathcal{D}_{\upsilon }\left( -\left(
1-i\right) HB\right) .  \label{App7}
\end{equation}%
Here, $D$\ is the parabolic cylinder function \cite{JCAM,PLA}, where
\begin{eqnarray}
\upsilon &=&-\frac{1}{2}+\frac{ig_{1}^{2}}{2}\sqrt{\frac{2E\mathcal{V}}{%
g_{2}^{2}}},  \label{App8} \\
\bar{\upsilon} &=&-\frac{1}{2}-\frac{ig_{1}^{2}}{2}\sqrt{\frac{2E\mathcal{V}%
}{g_{2}^{2}}},  \notag \\
B &=&\left( 2E\mathcal{V}g_{2}^{2}\right) ^{\frac{1}{4}}.  \notag
\end{eqnarray}%
We recall that $\left\vert z\right\vert ^{2}=z\bar{z}=\left( x+iy\right)
\left( x-iy\right) =x^{2}+y^{2}$. From Eqs. (\ref{App7}) we have%
\begin{equation}
\left\vert \upsilon \right\vert ^{2}=\upsilon \bar{\upsilon}=\left( -\frac{1%
}{2}\right) ^{2}+\frac{g_{1}^{4}}{4}\frac{2E\mathcal{V}}{g_{2}^{2}},
\end{equation}%
\begin{equation}
\frac{g_{1}^{4}}{4}\frac{2E\mathcal{V}}{g_{2}^{2}}=\left\vert \upsilon
\right\vert ^{2}-\frac{1}{4},
\end{equation}%
\begin{equation}
E=\frac{2g_{2}^{2}}{\mathcal{V}g_{1}^{4}}\left( \left\vert \upsilon
\right\vert ^{2}-\frac{1}{4}\right) .
\end{equation}%
From Eqs. (\ref{b3})-(\ref{bb8}) we have $\frac{g_{2}^{2}}{g_{1}^{4}}=\frac{%
\alpha _{2}}{6}$. This leads to Eq. (\ref{sc8}):%
\begin{equation}
E=\frac{\alpha _{2}}{3\mathcal{V}}\left( \left\vert \upsilon \right\vert
^{2}-\frac{1}{4}\right) .
\end{equation}

\end{document}